\documentclass{article}
\usepackage{spconf,amsmath,amssymb,graphicx,multirow}

\title{a unified multichannel far-field speech recognition system: combining neural beamforming with attention based end-to-end model}
\name{Dongdi Zhao$^{\star }$,  Jianbo Ma$^{\star}$, Lu Lu, Jinke Li, Xuan Ji, Lei Zhu, Fuming Fang, Ming Liu, Feijun Jiang\thanks{$^{\star }$ are first authors in alphabetical order and contributed equally. Work has been done in Alibaba Group.}}
\address{ Alibaba Group }
\begin{document}
%
\maketitle
\begin{abstract}
Far-field speech recognition is a challenging task that conventionally uses signal processing beamforming to attack noise and interference problem. But the performance has been found usually limited due to heavy reliance on environmental assumption. In this paper, we propose a unified multichannel far-field speech recognition system that combines the neural beamforming and transformer-based Listen, Spell, Attend (LAS) speech recognition system, which extends the end-to-end speech recognition system further to include speech enhancement. Such framework is then jointly trained to optimize the final objective of interest. Specifically, factored complex linear projection (fCLP) has been adopted to form the neural beamforming. Several pooling strategies to combine look directions are then compared in order to find the optimal approach. Moreover, information of the source direction is also integrated in the beamforming to explore the usefulness of source direction as a prior, which is usually available especially in multi-modality scenario. Experiments on different microphone array geometry are conducted to evaluate the robustness against spacing variance of microphone array. Large in-house databases are used to evaluate the effectiveness of the proposed framework and the proposed method achieve 19.26\% improvement when compared with a strong baseline.
\end{abstract}
\section{Introduction}
\label{sec:intro}

Automatic Speech Recognition (ASR) systems are conventionally composed of several components, such as deep neural network (DNN) based acoustic model and n-gram language models \cite{hinton2012deep}, which is referred as HMM/DNN. A limitation of this hybrid system is that it lacks of an objective that is shared by all the components in one system \cite{graves2014towards}. End-to-end automatic speech recognition (ASR) system then emerges as a competitor and enjoys popularity recent years mainly due to its simplicity and decent training process \cite{graves2014towards} \cite{chan2016listen}. Amongst those systems, Connectionist Temporal Classification (CTC) \cite{graves2006connectionist}\cite{graves2014towards}\cite{hannun2014deep}\cite{miao2015eesen} based and attention-based encoder-decoder  \cite{chan2016listen}\cite{chorowski2014end}\cite{bahdanau2014neural} are arguably the two most popular systems that has gained widely acceptance. 

Nevertheless, both of hybrids and end-to-end fashion systems, typically take spectrogram-based  features (e.g. filter-bank feature)  as inputs. Before spectrogram, a signal processing module is commonly served as the front-end processing part as to perform the tasks such as speech enhancement \cite{benesty2005speech}. This module is especially necessary when deal with far-field scenarios, for example, smart speakers where multichannel array techniques are usually employed. Several components, including acoustic echo cancellation (AEC), direction of arrival (DOA) and beamforming, are cascaded to form the entire module with each one targeting a particular task. Specifically, the technique of beamforming is usually accomplished by methods such as delay-and-sum, the Minimum Variance Distortionless Response (MVDR)  \cite{van2004optimum} \cite{ochiai2017multichannel},  and global sidelobe cancellation (GSC) \cite{buckley1986adaptive}, by exploiting signals from different microphones to find target speech source and suppressing interference and noise.

However, in real-world scenarios, the performance of the traditional beamforming techniques is often limited. This is partly due to the fact that these methods often make assumption of environments, such as stationary signal. As an alternative, neural beamforming technology has been proposed in conjunction with the traditional hybrid ASR system \cite{sainath2015speaker} \cite{sainath2016factored} with the possibility of making weaker assumption of the environment. For example, in \cite{heymann2016neural}, spectral masks are generated by neural networks to estimate power spectral density matrices to compute the beamforming coefficients. But it uses a separate objective like MVDR. In \cite{sainath2015speaker}, multichannel raw waveforms are trained directly on the convolutional, long short-term memory, deep neural network (CLDNN) \cite{sainath2015convolutional} and find it has similar effect as delay-and-sum in traditional beamforming, which drives the work in \cite{sainath2016factored} to separate spatial filtering and spectral filtering layer. But they are performed in time domain, which is computationally inefficient. In order to solve this problem, as a subsequent work, the factored complex linear projection (fCLP) in \cite{sainath2016factored}\cite{lu2020listen} has shown promising results with real-world smart speakers with much less computational burden. But the factored system lacks of an objective that is optimized by the entire system, while the holistic optimization may be better than prior knowledge \cite{graves2006connectionist}. 

Moreover, audio-visual speech enhancement has shown that visual information can offer extra source direction information \cite{michelsanti2021overview} and complements audio information only scenario. In \cite{xu2020neural}, target DOA inferred by visual algorithms has been merged into complex mask network, which is then fed into a MVDR beamformer. The results has been shown that this source direction information as a prior serves as an important feature for the beamformer. The source direction as a prior for the fCLP based neural beamformer has not been tested yet.

In order to tackle the aforementioned problem, in this work, we aim to combine complementary components of two different stages - neural beamforming and end-to-end speech recognition sub-system together such that the system optimizes a final objective. Several pooling strategies to combine look directions are then compared in order to find the optimal approach. Information of the source direction is also integrated in the beamforming to explore the usefulness of source direction as a prior. The attention based transformer LAS framework has been adopted as the back-end part. During the training phase, these two sub-systems are jointly trained. Large in-house databases are used to evaluate effectiveness of the proposed framework and the proposed method achieves 19.26\% improvement when compared with a strong baseline.

\section{CTC/attention system}
\label{sec:CTC/LAS}

\begin{figure}[htb]

  \centering
  \centerline{\includegraphics[width=6cm]{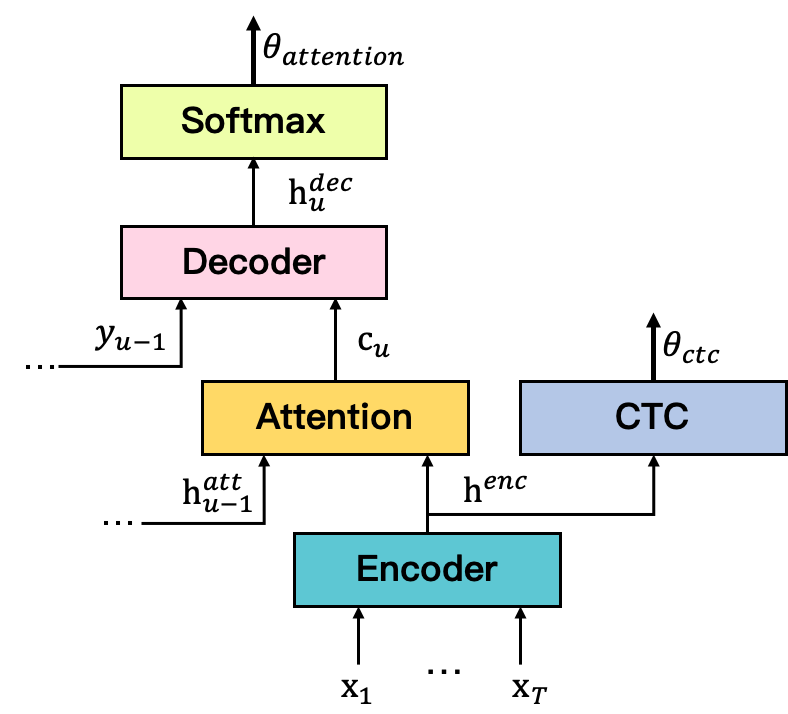}}
\caption{Attention-based End-to-End Architecture.}
\label{fig:las}
\end{figure}

LAS model consists of two sub-modules, namely encoder and decoder, as shown in Fig.\ref{fig:las}. The key operation of encoder is \texttt{Listen} which transforms acoustic features $\textbf{x}=(x_1,\cdots,x_T)$ into a high level representation $\textbf{h}=(h_1,\cdots,h_U)$ with $U{\leq}T$ as if down-sampling is applied. The key operation of decoder is \texttt{AttendAndSpell}. The attention mechanism integrates the encoder output $\textbf{h}$ and produces a context vector $c_u$ based on the previous decoder state $h^{att}_{u-1}$. The \texttt{Spell} function emits characters or words conditioned on $c_u$ and previous output labels $y_{1:u-1}$:
\begin{equation}\label{eq0-0}
P(\textbf{y}|\textbf{x})=\prod_{u}P(y_u|\textbf{x},y_{1:u-1}).
\end{equation}

\noindent Then the loss function of the model is computed as
\begin{equation}\label{eq0-1}
\theta_{attention}=-\mathrm{ln}P(\textbf{y}^{*}|\textbf{x})=-\sum_{u}\mathrm{ln}P(y_{u}^{*}|\textbf{x},y^{*}_{1:u-1})
\end{equation} 
where $\textbf{y}^{*}$ is the ground truth label.

Self-attention layer also has the advantage of parallel computation at all positions and position encodings are added to the acoustic features to inject relative position information. The attention mechanism used in the Transformer is multi-head attention, where each head computes Scaled Dot-Product Attention as

\begin{equation}\label{eq0-2}
\mathrm{Attention}(Q,K,V)=\mathrm{softmax}(\frac{QK^{T}}{\sqrt{d_k}})V.
\end{equation}

\noindent where $K$ and $V$ are the representations from all positions and $Q$ are the query vectors.

As in \cite{kim2017joint}, we add CTC loss function as an auxiliary task for encoder. The auxiliary task enforces the encoder to learn alignments between acoustic features and labels which results in faster convergence speed. The final objective function of the multi-task training framework is 
\begin{equation}\label{eq1}
\theta=\lambda \theta_{ctc}+(1-\lambda) \theta_{attention}
\end{equation} 

\noindent where $\lambda$ is a tunable parameter. 

\section{Proposed framework}
\label{sec:neuralBbeamformingASR}

The proposed architecture is shown as in Fig.\ref{fig:nnbfE2e}. This is a unified system that combines the neural beamforming and LAS speech recognition system. The details are described in the following sub-sections.

\subsection{Neural Beamforming}
\label{ssec:neuralBeamforming}

\begin{figure}[tb]

  \centering
  \centerline{\includegraphics[width=7cm]{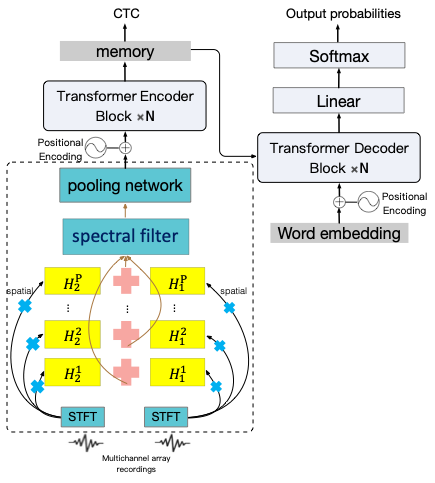}}

\caption{Architecture of proposed unified neural beamforming end-to-end speech recognition system. }
\label{fig:nnbfE2e}
\end{figure}

Illustrated as in Fig.\ref{fig:nnbfE2e}, the neural beamforming block contains two parts - the spatial filtering and spectral filtering components. The spatial filtering part is similar as to form different look directions, while spectral filtering is analog to find the most informative direction and extract features for recognition task.

Specifically, the input of this block is the multichannel waveforms, which are then transformed into frequency domain by short-time Fourier transform (STFT) with window size as $L$. The outcome is denoted as $X_c[t]\in \mathbb{C}^{K} $, where $t$ and $c$ are time and channel indices,  $K=N/2+1$, where $N$ is the number of fast Fourier transform (FFT). Suppose $P$ look directions are applied, then $P$ spatial filter $H_c^p\in \mathbb{C}^{K}$ will be applied to beamform the frequency domain signals as 
\begin{equation}\label{eq1}
Y_p[t]=\sum_{c=1}^{C}  X_c[t] \cdot H_c^p   
\end{equation}
where $\cdot$ denotes the matrices element-wise product.

As for the spectral filtering part, we use the factored Complex Linear Projection (fCLP) \cite{sainath2016reducing} as it reduces computational complexity and results in similar performance compared to a single channel system \cite{variani2016complex}. This is accomplished by using complex filterbank $S_f^p\in \mathbb{C}^K$ to spatial filter's outcome as 
\begin{equation}\label{eq2}
Z_f^p[t]=Y_p[t] \cdot S_f^p  
\end{equation}
The outputs of the filterbank is then summed and log compressed for each direction and spectral filter as 
\begin{equation}\label{eq3}
O_f^p[t]=\log |\sum_{k=1}^{K} Z_f^p[t, k]|
\end{equation}

\subsection{Neural Beamforming with ASR}
\label{ssec:neuralbeamformingASR}
\begin{figure}[tb]

\begin{minipage}[b]{0.3\linewidth}
  \centering
  \centerline{\includegraphics[width=1.8cm]{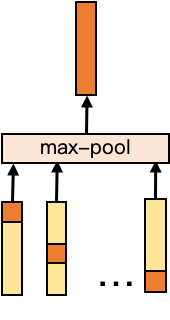}}
  \centerline{(a) Max-pool}\medskip
\end{minipage}
\begin{minipage}[b]{.3\linewidth}
  \centering
  \centerline{\includegraphics[width=1.8cm]{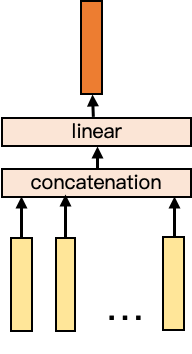}}
  \centerline{(b) Projection}\medskip
\end{minipage}
\hfill
\begin{minipage}[b]{0.37\linewidth}
  \centering
  \centerline{\includegraphics[width=2.4cm]{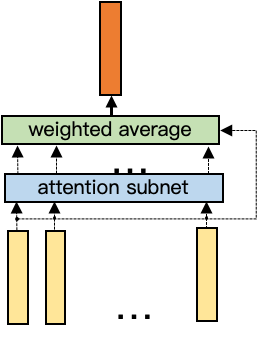}}
  \centerline{(c) Attention}\medskip
\end{minipage}
\caption{Illustration of different pooling methods.}
\label{fig:poolStrategy}
%
\end{figure}

The last step of the neural beamforming block serves as feature extraction when compared with the conventional way, e.g. filter-bank acoustic features. As can be seen in (\ref{eq3}), although elements in one spectral filter has been summed, there are $P$ directions that may be redundant. In the conventional signal processing module,  there are usually $P_{dsp}$ (e.g. $P_{dsp}=5$) look directions in the beamforming that is formed by algorithms like GSC. DOA is then used to choose the best matching direction of the beamforming as the output. Similarly, pooling layer over the look direction is then applied on $O_f^p[t]$ before fed into the encoder. In the proposed framework, we tried three different pooling strategies, as shown in Fig.\ref{fig:poolStrategy}. (1) \textbf{Max-pool}: the most significant features from different look directions were integrated into a single feature vector. (2) \textbf{Projection}: the features from all look directions were concatenated, and a linear layer was then used to project the concatenated feature into 40 dimension. (3) \textbf{Attention}: features from all directions were integrated according to their importance to recognition, the importance weight was computed by a learnable matrix $W$ and the output is the weighted summation as 


\begin{equation}\label{eq4}
O[t]=\sum_{p}^{P}O_p[t]\cdot\mathrm{softmax}(WO_p[t]).
\end{equation}

The rest of system is the same as described in section 2. As the multiplication operation in (\ref{eq2}-\ref{eq3}) and the norm can be represented by real matrix multiplication \cite{variani2016complex}, gradients are represented in real value and back-propagation can be used to train the entire model. 

When compared to the conventional method, another advantage of this unified framework also comes from the computational complexity. In the conventional signal processing module, beamforming algorithms may require matrix inversion to get the solution. As there are $P_{dsp}$ (e.g. $P_{dsp}=5$) look directions, multi-time matrix inversion needs to be processed, which is computational intensive. In the proposed framework, computations are mainly from (\ref{eq2}-\ref{eq3}), but the matrices multiplication can be well parallelized by modern processors.

\subsection{Integrating source direction}
\label{ssec:integrating_source_direction}

\begin{figure}[tb]

\begin{minipage}[b]{0.48\linewidth}
  \centering
  \centerline{\includegraphics[width=3.2cm]{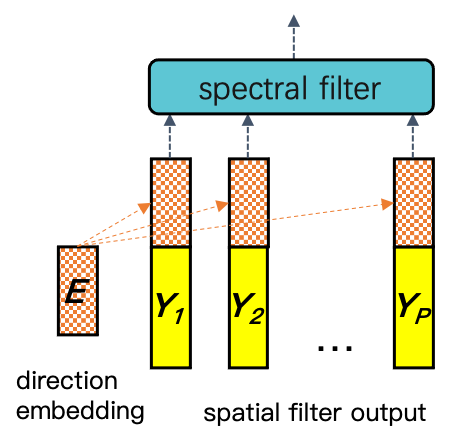}}
  \centerline{(a) Direction Aware Module}\medskip
\end{minipage}
\begin{minipage}[b]{.48\linewidth}
  \centering
  \centerline{\includegraphics[width=3.2cm]{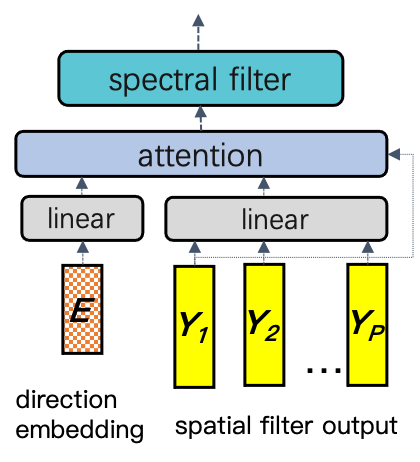}}
  \centerline{(b) Direction Attentive Module}\medskip
\end{minipage}
%
\caption{Illustration of integrating source direction.}
\label{fig:integrating_source_direction}
%
\end{figure}

As latest research indicates, incorporating source direction in beamforming may improve the speech separation performance \cite{xu2020neural}. The source direction as prior is usually available in scenario like multi-modality, in which source direction can be more accurately estimated by computer vision. In this work, the source direction information is integrated in aforementioned neural beamforming as illustrated in figure \ref{fig:integrating_source_direction}. Specifically, two different methods are explored. In both methods, source direction is the angle of source, which is then represented by angle embedding. Angle embeddings are learnable variables. First method augments neural beamforming as a direction aware module which concatenates angle embedding with outputs of each spatial filter in neural beamforming before projection as illustrated by figure \ref{fig:integrating_source_direction}.a. Second method applies a direction attentive module as illustrated by \ref{fig:integrating_source_direction}.b. This mechanism is similar as CLAS \cite{pundak2018deep}. The output of spatial filter is
\begin{equation}\label{eq4}
Y[t]=\sum_{p}^{P}\alpha_{p}(t) Y_p[t]
\end{equation}
where
\begin{equation}\label{eq4-1} 
u_{p}(t)=v^{T} \mathrm{tanh}(W_{p}^{0}Y_p+W_{p}^{e}E)
\end{equation}
\begin{equation}\label{eq4-2} 
\alpha_{p}(t)=\mathrm{softmax}(u_{p}(t))
\end{equation}
and $W_{p}^{0}$, $W_{p}^{e}$, $v$ are learnable variables. The idea is to use source direction in order to select the most informative direction.

\section{Experiments}
\label{sec:experiments}

We evaluated the proposed framework with large in-house databases. Specifically, speech recognition with conventional beamforming using GSC algorithm and neural beamforming were compared. Different strategies of pooling methods and integration of source direction in neural beamforming were evaluated. To assess the robustness, different spacing of microphone array are also included in the experiments.

\subsection{Databases}
\label{ssec:databases}

To fully test the effectiveness of the proposed framework, two different databases were used. The first one was collected in a near-field cellphone setting, which includes around 3000 hours data. The second is an in-house database which containing 2400 hours online audio data.

The near-field waveforms in the original databases were artificially corrupted by a room simulator, and then processed by adding different levels of noise and reverberation. Room simulator described in \cite{scheibler2018pyroomacoustics} was used. 
We randomly chose different room dimensions and the position of microphone array from a pools of samples. Room configurations with RT60s range from 50 to 500 ms. In this simulation task, we used two microphone arrays to generate two-channel recordings. The spacing of the microphone array was set as 4cm for training set. When adding noise, the level of noise is varying from utterance to another, with SNR obey to normal distribution from 0 to 20 dB. The distance between speech source and microphones varies in range of 0.5 to 7m. To test the robustness against the varying angle between speech source and interference, azimuths of speech and noise were randomly chosen from range of $\pm 180 $ degrees and elevation of speech and noise was constrained to be between [0.6m, 2.0m] and [0.4m, 3.0m]. We composed a evaluation set by randomly selecting utterances from the full-sized database. It roughly contains 15K utterance and is around 15h hours. Same simulation process was used as the training set.


For the in-house real data, online audio data are anonymiz-ed and extracted from on-line service. Compared to simulated data, real data is usually recorded in a less complicated environment with fewer noise sources and smaller room size which results in a higher SNR and lower reverberation. On the other hand, real data is more in-domain and has less generalization.

\subsection{System Configurations}
\label{sssec:configurations}

The baseline system is a transformer based CTC/Attention speech recognition system with conventional beamforming. GSC with five look directions and DOA was used to obtain the enhanced speech signal. Filter-bank features with 40 dimensions were then extracted before sent to the encoder. The inputs were optionally processed by offline weighted prediction error (WPE) for de-reverberation \cite{yoshioka2012generalization}.  The number of look direction $P$ in neural beamforming was chosen as 10 and the number of filters in spectral filtering part is 40. 


In the encode-decoder configuration, 7 transformer blocks and 2 transformer block were used for encoder and decoder. $\lambda=0.1$ in (4) was used in the entire experiments. Adam optimizer was used to train the networks with initial learning rate as 0.001. The learning rate is decayed as in \cite{vaswani2017attention}. Similarly, we also use warm up step which is set as 4000. The entire networks were trained with 30 epochs for both conventional and proposed framework. When decoding, beam search algorithm was used. An external LSTM-based language model was used to rescore the decoding paths in the manner of shallow fusion \cite{toshniwal2018comparison}.

\subsection{Results }
\label{sssec:results}

First experiment was conducted to evaluate the proposed neural beamforming end-to-end ASR system (NBE2E) and compared it with the baseline single-channel system (DSPE2E). Word error rate (WER) of both systems on the evaluation set with and without WPE processing were reported. To evaluate different pooling methods in neural beamforming, we also reported the performance of different pooling strategies (max-pool/proj/attn) with or without WPE processing (w/o WPE and w/ WPE).


\begin{table}[htb]
\centering
 \begin{tabular}{c c c c } 
 \hline
 Dataset & System & w/o WPE & w/ WPE \\ [0.5ex] 
 \hline \hline
 \multirow{4}{*}{Simulated data} & DSPE2E & 31.62 & 30.89 \\ 
 & NBE2E max-pool & 26.58 & 26.32  \\
 & NBE2E attn & 28.20 & 26.52 \\
 & NBE2E proj & \textbf{26.28} & \textbf{24.94} \\  [1ex] 
 \hline
 \multirow{2}{*}{Real data} & DSPE2E & 5.89 & - \\
 & NBE2E & \textbf{5.40} & - \\
 \hline
\end{tabular}
\caption{Results (word error rate \%) comparison of baseline system and proposed unified multichannel end-to-end system with different pooling strategies in neural beamforming and with or without WPE. }
\label{table:pooling}
\end{table}

The results are presented in table \ref{table:pooling}. 
Firstly, the proposed system achived WER of 24.94\%, while the one of baseline is 30.89\%, which accounts for 19.26\% relative improvement with the unified system. This supports the idea that extending end-to-end ASR system to unify neural beamforming is beneficial and suggests a better option for far-field scenarios. Furthermore, amongst the three different pooling approaches, the projection is the best and outperformed the attention method by relative 6.31\%. This is perhaps because the projection approach preserves more information. It is also observed that, for both systems, when cascaded with WPE, better performances can be obtained. Relative 5.10\% and 2.31\% improvements are obtained for proposed system and baseline. This suggests that de-reverberation is beneficial for both systems. 

Further experiments on 2400h in-house data are used to test the effectiveness of the proposed unified ASR system in real world scenario and the results are presented in table \ref{table:pooling}. As  negligible reverberation can be observed from real data, experiments on WPE were not conducted. The WER of the conventional DSP with end-to-end ASR system is 5.89\%, while the unified ASR system achieved 5.40\%, accounting for 8.32\% relative improvement. 


\subsection{Performance on different microphone array spacing}
\label{sec:dimensions}

To evaluate how the proposed system and the baseline system designed for a specific microphone array geometry perform when generalizing to other microphone array geometries, the effects of different microphone array spacing was also evaluated in this section. 
In this experiment, models are the same as those of the previous section. The difference is that the array geometry in the evaluation set is categorized as 4cm, 6cm, 8cm and 10cm. In particular, we used the projection pooling method in the proposed method as it performs the best. The results are presented in table \ref{table:2}. 

\begin{table}[h!]
\centering
 \begin{tabular}{c c c } 
 \hline
 Mic Spacing & DSPE2E & NBE2E \\ [0.5ex] 
 \hline\hline
 4cm & 30.89 & 24.94 \\ 
 6cm & 30.67 & 25.10 \\
 8cm & 31.14 & 25.54 \\
 10cm & 32.03 & 25.74 \\[1ex] 
 \hline
\end{tabular}
\caption{Results (word error rate \%) comparison of baseline system and proposed unified multichannel end-to-end system with different array geometry. }
\label{table:2}
\end{table}

From table \ref{table:2}, it can be seen that the proposed system obtained better performance in 4cm category. This is reasonable as the models are trained with data from microphone array with 4cm. But it is also observed that the baseline system degrades more sharply and fluctuates more severely than the one of the unified end-to-end system. For example, when the distance of microphone array is 10cm, the baseline degraded to 32.03\%. 

This may be because that algorithms in conventional DSP module rely more on the prior knowledge and neural beamforming has better generalization ability for different microphone array geometry. 

\subsection{Performance of integrating source direction}

To test the effectiveness of integrating source direction as prior in the proposed network, in this experiment, the 2400 hours real data was simulated as randomly choosing the source direction in RIR. The source direction was recorded as prior for the model training phase and evaluation phase. Note that we did not add artificial noise in this setting. As usually the DOA algorithms are not that accurate, we randomly add a perturbation in the simulated source direction. It uniformly distributed between $-10$ to $10$ degrees.

\label{table:sourceDirection}
\begin{table}[h!]
\centering
 \begin{tabular}{c c c } 
 \hline
 System & WER & WER-reduction \\ [0.5ex] 
 \hline\hline
 NBE2E Base & 10.90 & -- \\ 
 NBE2E Dir-aware & 6.90 & 36.70\% \\
 NBE2E Dir-atten & 6.86 & 37.06\% \\[1ex] 
 \hline
\end{tabular}
\caption{Results (word error rate \%) comparison of with or without integrating source direction. }
\label{table:2}
\end{table}

The results are presented in table \ref{table:sourceDirection}. The baseline is the proposed vanilla unified ASR system (without source direction), which achieved 10.90\% WER in this task. NBE2E Dir-aware and NBE2E Dir-atten represent the proposed system augmented with a direction aware module and a direction attentive module as described in Section \ref{ssec:integrating_source_direction} respectively. 6.90\% and 6.86\% WER were achieved by the augmented models, accounting for 37.06\% relative improvement. It can be seen that both two integrating methods outperformed the vanilla unified neural beamforming ASR system, indicating the superiority of using source direction as prior in the proposed unified ASR system. This finding is useful especially for multi-modality scenario.

\subsection{Performance against DOA resolution}
\label{sec:largeScale}
The last experiment was conducted to compare the proposed unified multichannel end-to-end system with the baseline in a real-world scenario. It has been reported that the conventional beamformer and DOA may results in incorrect direction estimation of speech and interference source, as the resolution is limited \cite{hyder2010direction}. As for the neural beamforming method, there is less reliance on the module functioning as estimating angle. The angle information of multichannel recordings is jointly learned and exploited by spatial and spectral filtering block. 

\begin{figure}[tb]

  \centering
  \centerline{\includegraphics[width=7.5cm]{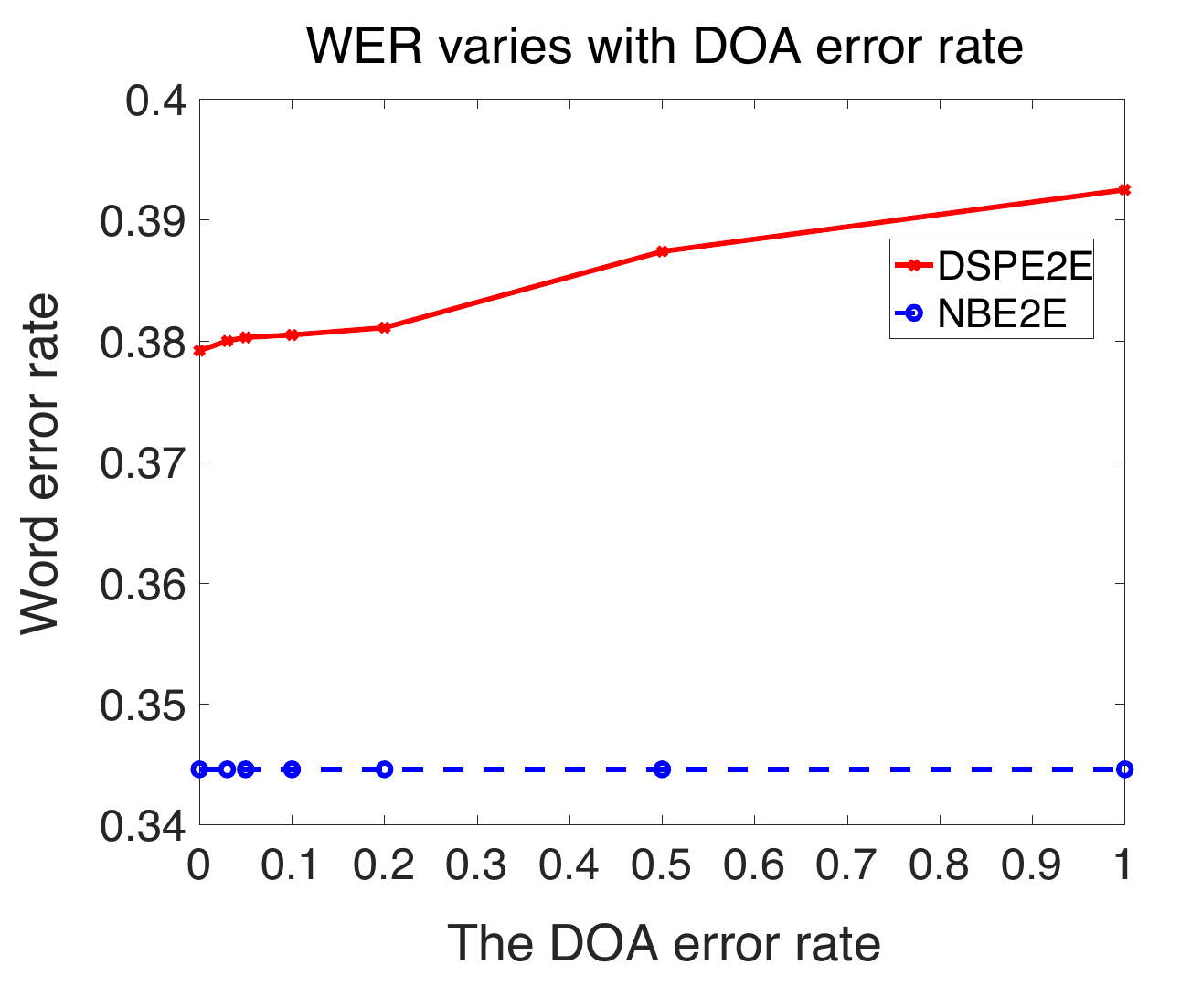}}
\caption{WER varies with DOA error rate.}
\label{fig:ad}
\end{figure}

Fig. \ref{fig:ad} gives the experimental results as the rate of incorrect direction estimation enlarging. We artificially controlled the incorrect rate as to give an idea how systems reacts to this factor. To simulate the resolution limitation, the gpuRIR toolkit is used to obtain evaluation data \cite{diaz2020gpurir}, which makes it different with the one of previous sections. As can be seen that the overall trend for conventional beamforming technique tends to degrade from 37.92\% to 39.25\% when the incorrect rate of angle estimate increases from 0 to 100\%. But since the proposed unified multichannel end-to-end system does not require angle estimation, the performance is flat, which indicates the proposed method could achieve more stable performance in the real-world scenario. 

\section{Conclusion}
\label{sec:conclusion}

In this work, we proposed a unified multichannel far-field speech recognition system that combines the neural beamforming and end-to-end speech recognition system. It combines the complementary components of two different stages - fCLP based neural beamforming and end-to-end speech recognition sub-system, together such that the system is robust against real-world scenarios with end-to-end fashion. Under this technique, several pooling strategies to combine looking directions are compared in order to find the optimal strategy. Information of the source direction is also integrated in the beamforming to explore the usefulness of source direction as a prior. The transformer LAS model has been adopted as the back-end part. These two sub-systems are jointly trained to optimize the final objective of recognition task. Large in-house databases are used to evaluate effectiveness of the proposed framework and the proposed method achieve 19.26\% improvement when compared with a strong baseline. We further augmented the proposed system with a direction integrating module and explored two integrating strategies. This brings significant gain and is especially useful in multi-modality scenario when source direction may be taken as prior.

\bibliographystyle{IEEEbib}
\bibliography{strings,refs}

\begin{thebibliography}{10}

\bibitem{hinton2012deep}
Geoffrey Hinton, Li~Deng, Dong Yu, George~E Dahl, Abdel-rahman Mohamed, Navdeep
  Jaitly, Andrew Senior, Vincent Vanhoucke, Patrick Nguyen, Tara~N Sainath,
  et~al.,
\newblock ``Deep neural networks for acoustic modeling in speech recognition:
  The shared views of four research groups,''
\newblock {\em IEEE Signal processing magazine}, vol. 29, no. 6, pp. 82--97,
  2012.

\bibitem{graves2014towards}
Alex Graves and Navdeep Jaitly,
\newblock ``Towards end-to-end speech recognition with recurrent neural
  networks,''
\newblock in {\em International conference on machine learning}, 2014, pp.
  1764--1772.

\bibitem{chan2016listen}
William Chan, Navdeep Jaitly, Quoc Le, and Oriol Vinyals,
\newblock ``Listen, attend and spell: A neural network for large vocabulary
  conversational speech recognition,''
\newblock in {\em 2016 IEEE International Conference on Acoustics, Speech and
  Signal Processing (ICASSP)}. IEEE, 2016, pp. 4960--4964.

\bibitem{graves2006connectionist}
Alex Graves, Santiago Fern{\'a}ndez, Faustino Gomez, and J{\"u}rgen
  Schmidhuber,
\newblock ``Connectionist temporal classification: labelling unsegmented
  sequence data with recurrent neural networks,''
\newblock in {\em Proceedings of the 23rd international conference on Machine
  learning}, 2006, pp. 369--376.

\bibitem{hannun2014deep}
Awni Hannun, Carl Case, Jared Casper, Bryan Catanzaro, Greg Diamos, Erich
  Elsen, Ryan Prenger, Sanjeev Satheesh, Shubho Sengupta, Adam Coates, et~al.,
\newblock ``Deep speech: Scaling up end-to-end speech recognition,''
\newblock {\em arXiv preprint arXiv:1412.5567}, 2014.

\bibitem{miao2015eesen}
Yajie Miao, Mohammad Gowayyed, and Florian Metze,
\newblock ``Eesen: End-to-end speech recognition using deep rnn models and
  wfst-based decoding,''
\newblock in {\em 2015 IEEE Workshop on Automatic Speech Recognition and
  Understanding (ASRU)}. IEEE, 2015, pp. 167--174.

\bibitem{chorowski2014end}
Jan Chorowski, Dzmitry Bahdanau, Kyunghyun Cho, and Yoshua Bengio,
\newblock ``End-to-end continuous speech recognition using attention-based
  recurrent nn: First results,''
\newblock {\em arXiv preprint arXiv:1412.1602}, 2014.

\bibitem{bahdanau2014neural}
Dzmitry Bahdanau, Kyunghyun Cho, and Yoshua Bengio,
\newblock ``Neural machine translation by jointly learning to align and
  translate,''
\newblock {\em arXiv preprint arXiv:1409.0473}, 2014.

\bibitem{benesty2005speech}
Jacob Benesty, Shoji Makino, and Jingdong Chen,
\newblock {\em Speech enhancement},
\newblock Springer Science \& Business Media, 2005.

\bibitem{van2004optimum}
Harry~L Van~Trees,
\newblock {\em Optimum array processing: Part IV of detection, estimation, and
  modulation theory},
\newblock John Wiley \& Sons, 2004.

\bibitem{ochiai2017multichannel}
Tsubasa Ochiai, Shinji Watanabe, Takaaki Hori, and John~R Hershey,
\newblock ``Multichannel end-to-end speech recognition,''
\newblock in {\em International Conference on Machine Learning}. PMLR, 2017,
  pp. 2632--2641.

\bibitem{buckley1986adaptive}
K~Buckley and L~Griffiths,
\newblock ``An adaptive generalized sidelobe canceller with derivative
  constraints,''
\newblock {\em IEEE Transactions on antennas and propagation}, vol. 34, no. 3,
  pp. 311--319, 1986.

\bibitem{sainath2015speaker}
Tara~N Sainath, Ron~J Weiss, Kevin~W Wilson, Arun Narayanan, Michiel Bacchiani,
  et~al.,
\newblock ``Speaker location and microphone spacing invariant acoustic modeling
  from raw multichannel waveforms,''
\newblock in {\em 2015 IEEE Workshop on Automatic Speech Recognition and
  Understanding (ASRU)}. IEEE, 2015, pp. 30--36.

\bibitem{sainath2016factored}
Tara~N Sainath, Ron~J Weiss, Kevin~W Wilson, Arun Narayanan, and Michiel
  Bacchiani,
\newblock ``Factored spatial and spectral multichannel raw waveform cldnns,''
\newblock in {\em 2016 IEEE International Conference on Acoustics, Speech and
  Signal Processing (ICASSP)}. IEEE, 2016, pp. 5075--5079.

\bibitem{heymann2016neural}
Jahn Heymann, Lukas Drude, and Reinhold Haeb-Umbach,
\newblock ``Neural network based spectral mask estimation for acoustic
  beamforming,''
\newblock in {\em 2016 IEEE International Conference on Acoustics, Speech and
  Signal Processing (ICASSP)}. IEEE, 2016, pp. 196--200.

\bibitem{sainath2015convolutional}
Tara~N Sainath, Oriol Vinyals, Andrew Senior, and Ha{\c{s}}im Sak,
\newblock ``Convolutional, long short-term memory, fully connected deep neural
  networks,''
\newblock in {\em 2015 IEEE International Conference on Acoustics, Speech and
  Signal Processing (ICASSP)}. IEEE, 2015, pp. 4580--4584.

\bibitem{lu2020listen}
He~Weipeng, Lu~Lu, Zhang Biqiao, Mahadeokar Jay, Kalgaonkar Kaustubh, and
  Fuegen Christian,
\newblock ``Spatial attention for far-field speech recognitionn with deep
  beamforming neural networks,''
\newblock in {\em 2020 IEEE International Conference on Acoustics, Speech and
  Signal Processing (ICASSP)}. IEEE, 2020.

\bibitem{michelsanti2021overview}
Daniel Michelsanti, Zheng-Hua Tan, Shi-Xiong Zhang, Yong Xu, Meng Yu, Dong Yu,
  and Jesper Jensen,
\newblock ``An overview of deep-learning-based audio-visual speech enhancement
  and separation,''
\newblock {\em IEEE/ACM Transactions on Audio, Speech, and Language
  Processing}, 2021.

\bibitem{xu2020neural}
Yong Xu, Meng Yu, Shi-Xiong Zhang, Lianwu Chen, Chao Weng, Jianming Liu, and
  Dong Yu,
\newblock ``Neural spatio-temporal beamformer for target speech separation,''
\newblock {\em arXiv preprint arXiv:2005.03889}, 2020.

\bibitem{kim2017joint}
Suyoun Kim, Takaaki Hori, and Shinji Watanabe,
\newblock ``Joint ctc-attention based end-to-end speech recognition using
  multi-task learning,''
\newblock in {\em 2017 IEEE international conference on acoustics, speech and
  signal processing (ICASSP)}. IEEE, 2017, pp. 4835--4839.

\bibitem{sainath2016reducing}
Tara~N Sainath, Arun Narayanan, Ron~J Weiss, Ehsan Variani, Kevin~W Wilson,
  Michiel Bacchiani, and Izhak Shafran,
\newblock ``Reducing the computational complexity of multimicrophone acoustic
  models with integrated feature extraction,''
\newblock 2016.

\bibitem{variani2016complex}
Ehsan Variani, Tara~N Sainath, Izhak Shafran, and Michiel Bacchiani,
\newblock ``Complex linear projection (clp): A discriminative approach to joint
  feature extraction and acoustic modeling,''
\newblock 2016.

\bibitem{pundak2018deep}
Golan Pundak, Tara~N Sainath, Rohit Prabhavalkar, Anjuli Kannan, and Ding Zhao,
\newblock ``Deep context: end-to-end contextual speech recognition,''
\newblock in {\em 2018 IEEE spoken language technology workshop (SLT)}. IEEE,
  2018, pp. 418--425.

\bibitem{scheibler2018pyroomacoustics}
Robin Scheibler, Eric Bezzam, and Ivan Dokmani{\'c},
\newblock ``Pyroomacoustics: A python package for audio room simulation and
  array processing algorithms,''
\newblock in {\em 2018 IEEE International Conference on Acoustics, Speech and
  Signal Processing (ICASSP)}. IEEE, 2018, pp. 351--355.

\bibitem{yoshioka2012generalization}
Takuya Yoshioka and Tomohiro Nakatani,
\newblock ``Generalization of multi-channel linear prediction methods for blind
  mimo impulse response shortening,''
\newblock {\em IEEE Transactions on Audio, Speech, and Language Processing},
  vol. 20, no. 10, pp. 2707--2720, 2012.

\bibitem{vaswani2017attention}
Ashish Vaswani, Noam Shazeer, Niki Parmar, Jakob Uszkoreit, Llion Jones,
  Aidan~N Gomez, {\L}ukasz Kaiser, and Illia Polosukhin,
\newblock ``Attention is all you need,''
\newblock in {\em Advances in neural information processing systems}, 2017, pp.
  5998--6008.

\bibitem{toshniwal2018comparison}
Shubham Toshniwal, Anjuli Kannan, Chung-Cheng Chiu, Yonghui Wu, Tara~N Sainath,
  and Karen Livescu,
\newblock ``A comparison of techniques for language model integration in
  encoder-decoder speech recognition,''
\newblock in {\em 2018 IEEE spoken language technology workshop (SLT)}. IEEE,
  2018, pp. 369--375.

\bibitem{hyder2010direction}
Md~Mashud Hyder and Kaushik Mahata,
\newblock ``Direction-of-arrival estimation using a mixed $\ell _ {2, 0} $ norm
  approximation,''
\newblock {\em IEEE Transactions on Signal processing}, vol. 58, no. 9, pp.
  4646--4655, 2010.

\bibitem{diaz2020gpurir}
David Diaz-Guerra, Antonio Miguel, and Jose~R Beltran,
\newblock ``gpurir: A python library for room impulse response simulation with
  gpu acceleration,''
\newblock {\em Multimedia Tools and Applications}, pp. 1--19, 2020.

\end{thebibliography}

\end{document}